\documentclass[a4paper]{jpconf}
\usepackage{graphicx}
\usepackage{comment}
\begin{document}
\title{One-loop on-shell and  
off-shell decay $H^*\rightarrow VV$ 
at future $e^-e^-$ collider}
\author{Anh Thu Nguyen$^{(a,b)}$, 
Dzung Tri Tran$^{(a,b)}$ 
and Khiem Hong Phan$^{(c,d)}$}
\address{
$^{a)}$University of Science, 
Ho Chi Minh City $700000$, Vietnam\\
$^{b)}$Vietnam National University, 
Ho Chi Minh City $700000$, Vietnam\\
$^{c)}$Institute of Fundamental and Applied Sciences, 
Duy Tan University, Ho Chi Minh City $700000$, Vietnam\\
$^{d)}$Faculty of Natural Sciences, Duy Tan University, 
Da Nang City $550000$, Vietnam
}
\ead{phanhongkhiem@duytan.edu.vn, nathu@hcmus.edu.vn}
\begin{abstract}
One-loop on-shell and 
off-shell decays $H\rightarrow VV$ 
with $VV=\gamma\gamma, Z\gamma, ZZ$ are 
presented in this 
paper. The effects of one-loop on-shell and 
off-shell Higgs decays in 
Higgs productions at $e^-e^-$ 
collisions are also then examined. We find 
that the impacts of one-loop Higgs decays 
are significant and they are must be taken 
into account at $e^-e^-$ collision.
\end{abstract}
\section{Introduction}
After discovering Standard Model-like 
Higgs boson~\cite{ATLAS:2012yve,CMS:2012qbp}, 
the precise measurements for Higgs productions
and Higgs decay channels are one of main 
targets at the High Luminosity Large Hadron 
Collider (HL-LHC) and future lepton colliders 
(LC). Most of the on-shell Higgs decay 
processes are probed at the LHC. The data 
are in agreement with Standard Model 
(SM) predictions. In order to study the Higgs 
sector at different energy scales, 
off-shell Higgs decay channels are 
considerable interests at future colliders. 
Recently, off-shell 
Higgs decay $H^*\rightarrow Z^*Z^* \rightarrow 4$ 
leptons have been measured at the LHC in 
\cite{CMS:2014quz,
ATLAS:2015cuo,ATLAS:2015pre,ATLAS:2018jym,CMS:2019ekd,
CMS:2021ziv}. We also can find many
interesting points from the phenomenological 
studies for off-shell Higgs decays in 
\cite{Lee:2018fxj,Goncalves:2020vyn,Goncalves:2017iub,
Haisch:2021hvy,
Azatov:2022kbs,
Cacciapaglia:2014rla,Logan:2014ppa,Englert:2014ffa,
Chen:2015iha,
Goncalves:2018pkt,Dwivedi:2016xwm}. In all mentioned
references, the authors have not only provided 
the framework for testing the Higgs scalar 
sector at different energy scales but also 
discussed the possibility to explore new physics 
beyond the SM (BSM).

From theoretical computations, 
the evaluations for 
one-loop and two-loop QCD corrections
to the off-shell Higgs decay 
$H^*\rightarrow ZZ$  in both signals 
and backgrounds at the LHC 
have been found in Ref.~\cite{Caola:2015ila, 
Caola:2015psa,Campbell:2016ivq,Caola:2016trd,Grober:2019kuf,
Davies:2020lpf,Alioli:2021wpn,Grazzini:2021iae,Buonocore:2021fnj,
Haisch:2022rkm}. While one-loop electroweak corrections to 
Higgs boson decay into $\gamma\gamma, Z\gamma, ZZ$ have been calculated in 
Refs.~\cite{Marciano:2011gm,Phan:2020xih,
Hue:2017cph,Phan:2021pcc,
Kniehl:1990mq, Pierce:1992hg, Hollik:2011xd,
Boselli:2015aha,Bredenstein:2006rh,
Bredenstein:2006ha}. 
In this paper, we present 
one-loop on-shell and 
off-shell decays $H\rightarrow VV$ 
with $VV=\gamma\gamma, Z\gamma, ZZ$. 
The effects of one-loop on-shell and 
off-shell Higgs decays in
Higgs productions at $e^-e^-$ future 
lepton colliders are then examined. We find 
that the impacts of one-loop Higgs decays 
are significant and they are must be taken 
into account at $e^-e^-$ collision.

The layout of the paper is as follows: in section $2$, 
we discuss the calculation for one-loop
Higgs boson decay into $VV$. We then study
the impacts
of one-loop Higgs decays to $VV$
in Higgs productions at $e^-e^-$ collision. 
All signal productions 
$e^-e^- \rightarrow e^-e^- H^* \rightarrow e^-e^- (VV)$ are 
examined. Conclusions and outlooks 
are shown in section $3$. 
\section{Calculations}
In this section, the calculations for 
one-loop decay channels $H^* \rightarrow ZZ$  
within the Standard Model framework are discussed.
The computations are performed 
in 't Hooft-Veltman (or $R_{\xi = 1}$)
gauge. 
The decay process $H^* \rightarrow ZZ$ consists
of three group Feynman diagrams. In the first
group, we have all fermions exchanging in the loop.
With $W$-bosons and charged goldstone bosons 
and ghost particles exchanging in one-loop 
diagrams, we have group $2$. In the group 
$3$, one has all $Z$- or $H$-boson and 
neutral goldstone bosons propagating in 
the loop. For cancelling ultraviolet 
divergent (UV-divergent), we consider the 
counter-term diagram for vertex $HZZ$ (as
plotted in group $0$).
All Feynman one-loop diagrams are 
plotted in the appendix. One-loop 
amplitudes for $H^* \rightarrow 
\gamma \gamma, Z \gamma$ can be 
derived directly from the results of 
$H^* \rightarrow ZZ$.

The general one-loop amplitude for one-shell
and off-shell
Higgs decays into two on-shell vector bosons 
$H^*(p_H) \rightarrow
V_\mu (q_1) V_\nu (q_2)$ can be expressed
in terms of Lorentz structure 
as follows:
\begin{eqnarray}
\mathcal{A}_{H^*\rightarrow VV}
&=&
\Bigg[
F_{00}^{VV} 
\,
g^{\mu \nu}
+ F_{21}^{VV} 
\,
q_2^\mu q_1^\nu 
\Bigg]
\varepsilon^*_\mu (q_1) 
\varepsilon^*_\nu (q_2). 
\end{eqnarray}
Where $\varepsilon^*_\mu (q_1), \; 
\varepsilon^*_\nu (q_2)$ are polarization 
vectors of two external vector bosons.
The scalar functions
$F_{00}^{VV}=F_{00}^{VV} (M_{VV}^2,M_{V_1}^2,M_{V_2}^2)$
and $F_{21}^{VV}=F_{21}^{VV} (M_{VV}^2,M_{V_1}^2,M_{V_2}^2)$
are so-called one-loop form factors. 
They are functions
of $p_H^2=M_{VV}^2, M_{V_1}^2, M_{V_2}^2$. 
In the current work, we only focus on
the case of two real
external vector bosons in final state. 
Therefore, we have only the form factors 
$F_{00}^{VV}$ and $F_{21}^{VV}$ contributing
to the decay rates. All related kinematic
variables are given by:
\begin{eqnarray}
p_H^2=(q_1+q_2)^2=M_{VV}^2, \quad
q_1^2 = M_{V_1}^2,\quad q_2^2 = M_{V_2}^2,\quad
2(q_1 \cdot q_2) =
M_{VV}^2 - M_{V_1}^2 - M_{V_2}^2.
\end{eqnarray}

All Feynman amplitudes for $H^* \rightarrow ZZ$ 
are first written down. One then performs all
Dirac traces and Lorentz contractions in $d$ 
dimensions with the help of 
{\tt Package-X}~\cite{Patel:2015tea}. 
Subsequently, the amplitudes are then expressed 
in terms of tensor one-loop integrals. 
Following tensor reduction in~\cite{Denner:2005nn}, 
all tensor one-loop integrals are then 
decomposed into scalar one-loop Passarino-Veltman 
(PV-functions) functions. The PV-functions
can be computed numerically 
by using {\tt LoopTools}~\cite{Hahn:1998yk}.
As a result, the decay rates can be evaluated
numerically. Analytic expressions for 
all one-loop form factors of Higgs 
decay to $\gamma\gamma, Z\gamma, ZZ$ channels
can be found in \cite{Phan:2020xih,Phan:2021pcc,phkhiem}. 

General one-loop decay rate formulas for 
Higgs decays into two on-shell vector 
bosons $V_\mu (q_1) V_\nu (q_2)$ are 
derived as 
follows:
\begin{eqnarray}
\Gamma_{H^*\rightarrow VV}
&=&
\frac{(2 \pi)^4}{2 M_{VV}}
\int
\,
d\Phi_2 (M_{VV}^2,M_{V_1}^2,M_{V_2}^2)
\times
\sum \limits_{\textrm{pol} }
|\mathcal{A}_{H^*\rightarrow VV}|^2.
\end{eqnarray}
Where the phase space of 
$1 \rightarrow 2$ is given by
\begin{eqnarray}
\int d\Phi_2(M_{VV}^2,M_{V_1}^2,M_{V_2}^2)
&=&
\frac{\sqrt{\lambda(M_{VV}^2,M_{V_1}^2,M_{V_2}^2)}
}{128 \pi^5\; M_{VV}^2}.
\end{eqnarray}
Here we use Kall\"en function which is 
defined by $\lambda(a,b,c) = (a-b-c)^2 - 4bc$.

In order to calculate the squared amplitude 
for $H^* \rightarrow V V$, we first parameterize
the sum of polarization vectors for two 
final bosons in general form as follows:
\begin{eqnarray}
\sum \limits_{\textrm{pol}}
\varepsilon_\alpha (q_i)
\varepsilon^*_\beta (q_i)  
&=&
- g_{\alpha \beta}
+
\delta_{V_i} 
\frac{q_{i,\alpha} q_{i,\beta}}{M_{V_i}^2}.
\end{eqnarray}
Here we take $\delta_{V_i} = 0, 1$ for 
photon $\gamma$ and $Z$-boson, respectively.
Thus, the squared amplitude is casted into 
form of
\begin{eqnarray}
\sum \limits_{\textrm{pol}}
|\mathcal{A}_{H^*\rightarrow VV}|^2
&=&
\frac{1}{
16 M_{V_1}^2 M_{V_2}^2}
\Bigg\{
C_{00}\Big|F_{00}^{VV} (M_{VV}^2,M_{V_1}^2,M_{V_2}^2)\Big|^2 
+ 
C_{21}\Big|F_{21}^{VV} (M_{VV}^2,M_{V_1}^2,M_{V_2}^2)\Big|^2
\nonumber \\
&&\hspace{2cm}
+
C_{0021}\mathcal{R}e\Big[F_{00}^{VV}(M_{VV}^2,M_{V_1}^2,M_{V_2}^2)
F_{21}^{VV, *}(M_{VV}^2,M_{V_1}^2,M_{V_2}^2) \Big]
\Bigg\}. \nonumber\\
\end{eqnarray}
Where all related coefficients 
are listed as follows:
\begin{eqnarray}
C_{00}&=&4 \delta_{V_1} 
\Bigg[
\delta_{V_2} 
\Big(M_{V_1}^2
+M_{V_2}^2
-M_{VV}^2\Big)^2
-4 M_{V_1}^2 M_{V_2}^2
\Bigg]
-16 M_{V_1}^2 M_{V_2}^2 (\delta_{V_2}-4), 
\\
C_{21} &=&
\delta_{V_1} 
\Big(M_{V_1}^2
+M_{V_2}^2
-M_{VV}^2\Big)^2 
\Bigg[
\delta_{V_2} 
\Big(M_{V_1}^2
+M_{V_2}^2
-M_{VV}^2\Big)^2
-4 M_{V_1}^2 M_{V_2}^2
\Bigg]
\nonumber \\
&&\hspace{0.0cm} 
-4 M_{V_1}^2 M_{V_2}^2 
\Bigg[
\delta_{V_2} 
\Big(M_{V_1}^2
+M_{V_2}^2
-M_{VV}^2\Big)^2
-4 M_{V_1}^2 M_{V_2}^2
\Bigg], \\
C_{0021}&=&
16 M_{V_1}^2 M_{V_2}^2 
\Big(M_{V_1}^2
+M_{V_2}^2
-M_{VV}^2
\Big) 
(\delta_{V_2}-1)
\nonumber \\
&&\hspace{0cm}
-4 \delta_{V_1} 
\Big(M_{V_1}^2
+M_{V_2}^2
-M_{VV}^2\Big) 
\Bigg[
\delta_{V_2} 
\Big(M_{V_1}^2
+M_{V_2}^2
-M_{VV}^2
\Big)^2
-4 M_{V_1}^2 M_{V_2}^2
\Bigg].
\end{eqnarray}
As a result, 
one-loop decay rates for Higgs decays 
into on-shell $\gamma \gamma, Z \gamma, ZZ$ 
are expressed explicitly as follows:
\begin{eqnarray}
\label{decaygg}
\Gamma_{H^*\rightarrow \gamma \gamma}
&=&
\frac{\sqrt{\lambda(M_{\gamma \gamma}^2,0,0)}}
{(16 \pi)  M_{\gamma \gamma}^3}
\Bigg\{
4 \Big|F_{00}^{\gamma \gamma} 
(M_{\gamma\gamma}^2,0,0)\Big|^2
- M_{\gamma \gamma}^4 
\Big|F_{21}^{\gamma \gamma} 
(M_{\gamma\gamma}^2,0,0) \Big|^2
\Bigg\},  \\
\label{decayZg}
\Gamma_{H^*\rightarrow Z \gamma}
&=&
\frac{\sqrt{\lambda(M_{Z \gamma}^2,M_Z^2,0)}}
{(64 \pi) M_{Z \gamma}^3}
\Bigg\{
12 \Big|F_{00}^{Z \gamma} (M_{Z\gamma}^2,M_Z^2,0)\Big|^2
\\
&&\hspace{4.0cm}
- \Big(M_Z^2-M_{Z \gamma}^2\Big)^2
\Big|F_{21}^{Z \gamma} (M_{Z\gamma}^2,M_Z^2,0)\Big|^2
\Bigg\}, 
\nonumber \\
\Gamma_{H^*\rightarrow Z Z}
&=&
\frac{\sqrt{\lambda(M_{ZZ}^2,M_Z^2,M_Z^2)}}
{(256 \pi)  M_Z^4 M_{ZZ}^3}
\times
\\
&&\hspace{0cm} \times
\Bigg\{
\Big(48 M_Z^4
-16 M_Z^2 M_{ZZ}^2
+4 M_{ZZ}^4\Big)
\Big|F_{00}^{ZZ} (M_{ZZ}^2,M_Z^2,M_Z^2)\Big|^2 
\nonumber \\
&&\hspace{0cm}
+ \Big(16 M_Z^4 M_{ZZ}^4
-8 M_Z^2 M_{ZZ}^6
+M_{ZZ}^8\Big)
\Big|F_{21}^{ZZ} (M_{ZZ}^2,M_Z^2,M_Z^2)\Big|^2
\nonumber \\
&&\hspace{0cm}
+ \Big(32 M_Z^4 M_{ZZ}^2
-24 M_Z^2 M_{ZZ}^4
+4 M_{ZZ}^6\Big)
\times
\nonumber \\
&&\hspace{1.3cm} \times
\mathcal{R}e\Big[F_{00}^{ZZ} (M_{ZZ}^2,M_Z^2,M_Z^2) 
\cdot
F_{21}^{ZZ, *} (M_{ZZ}^2,M_Z^2,M_Z^2)\Big]
\Bigg\}. 
\nonumber
\end{eqnarray}

In phenomenological results, we study 
the impacts of on-shell and 
off-shell Higgs decay rates
$H^* \rightarrow V V$ in Higgs productions
at $e^-e^-$ collision. The cross sections for 
the processes $e^- e^- \rightarrow 
e^- e^- H^* \rightarrow e^- e^- (V V)$ are computed.
Differential cross sections with respect to 
$M_{VV}$ are given~\cite{phkhiem}:
\begin{eqnarray}
\frac{
d\sigma^{e^- e^- \rightarrow e^- e^- (VV)} (\sqrt{s},M_{VV})
}{d M_{VV}}
&=&
\frac{(2 M_{VV})\; \sigma^{e^- e^- \rightarrow 
e^- e^- H^*} (\sqrt{s},M_{VV})
}{\Big[(M_{VV}^2-M_H^2)^2 + (\Gamma_H M_H)^2\Big]}
\frac{M_{VV} \, \Gamma_{H^*\rightarrow VV} (M_{VV})}{\pi}.
\nonumber\\
\end{eqnarray}
Where analytic formulas for cross sections
of $\sigma^{e^- e^- \rightarrow 
e^- e^- H^*}$ 
have reported in Ref.~\cite{Hikasa:1985ee}.
In this paper, differential cross sections 
with respect
to $M_{VV}$ for all the processes 
$e^-e^- \rightarrow e^-e^- \gamma\gamma, Z\gamma, ZZ$ are plotted at center-of-mass energies (CMS energies) $\sqrt{s} = 350, 500$ GeV.

\begin{figure}[ht]
\centering
$\begin{array}{cc}
\hspace*{-4.5cm}
d\sigma/dM_{\gamma\gamma} [\textrm{fb}]
& 
\hspace*{-4.5cm}
d\sigma/dM_{\gamma\gamma} [\textrm{fb}] \\
\includegraphics[width=7.7cm, height=7.5cm]
{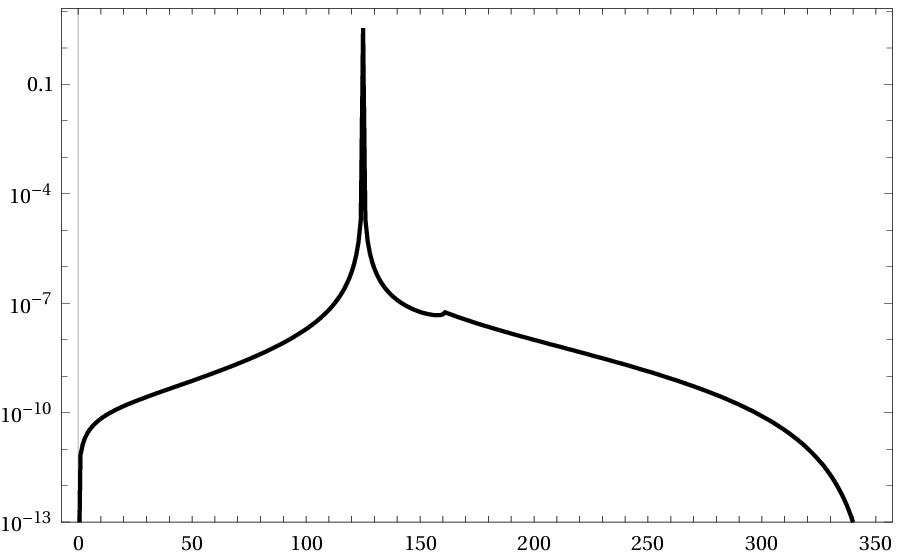}
&
\includegraphics[width=7.7cm, height=7.5cm]
{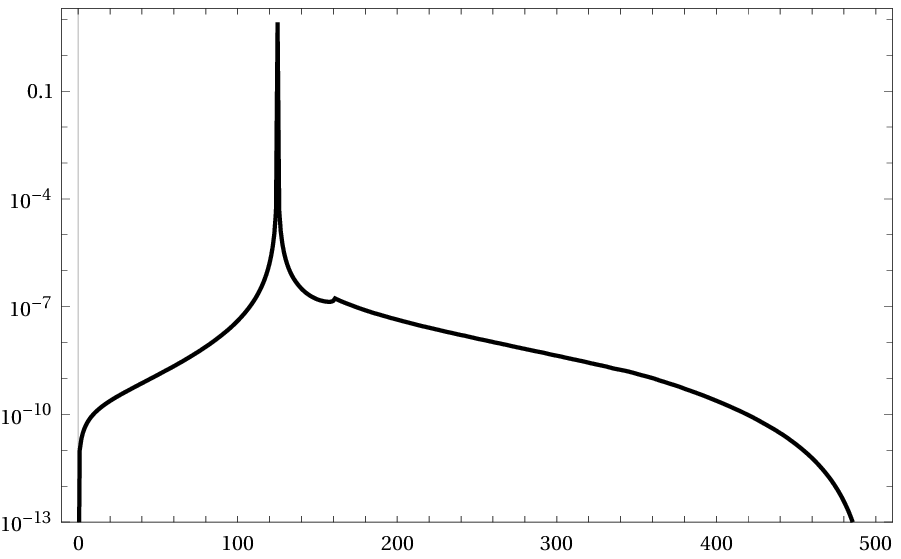}
\\
\hspace{6cm}
M_{\gamma\gamma} [\textrm{GeV}]  &
\hspace{6cm}
M_{\gamma\gamma} [\textrm{GeV}]
\end{array}$
\caption{\label{gammagamma} Differential 
cross sections with respect to $M_{\gamma\gamma}$ 
for process $e^- e^- \rightarrow 
e^- e^- H^* \rightarrow e^- e^- (\gamma\gamma)$ 
at $\sqrt{s} = 350$ GeV (left panel)
and $500$ GeV (right panel).}
\end{figure}

\begin{figure}[ht]
\centering
$\begin{array}{cc}
\hspace*{-4.5cm}
d\sigma/dM_{Z\gamma} [\textrm{fb}]
& 
\hspace*{-4.5cm}
d\sigma/dM_{Z\gamma} [\textrm{fb}] \\
\includegraphics[width=7.7cm, height=7.5cm]
{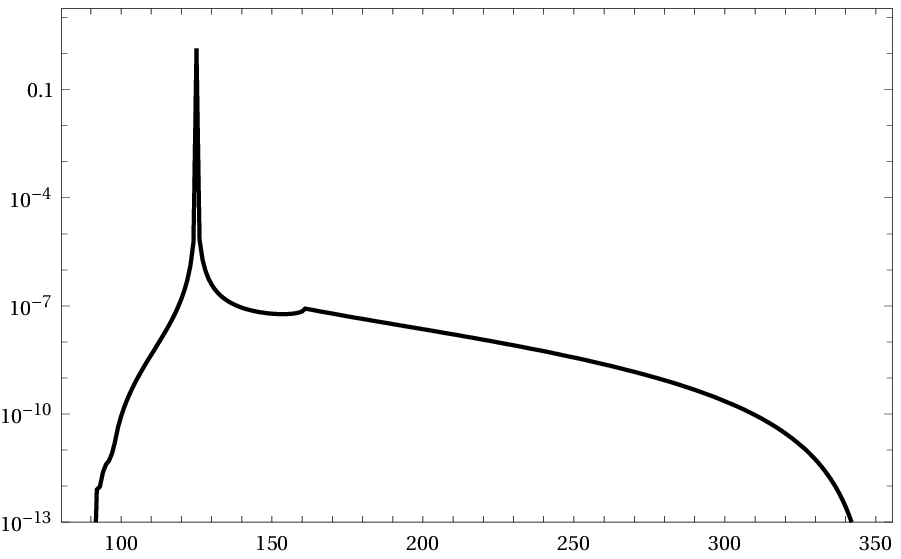}
&
\includegraphics[width=7.7cm, height=7.5cm]
{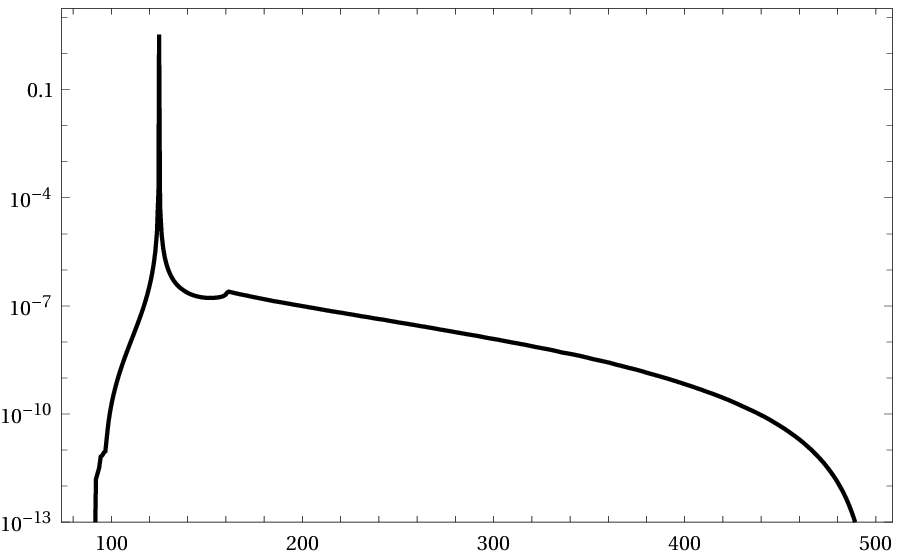}
\\
\hspace{6cm}
M_{Z\gamma} [\textrm{GeV}]  &
\hspace{6cm}
M_{Z\gamma} [\textrm{GeV}]
\end{array}$
\caption{\label{Zgamma} Differential cross 
sections with respect to $M_{Z\gamma}$ 
for process $e^- e^- \rightarrow 
e^- e^- H^* \rightarrow e^- e^- (Z\gamma)$ 
at $\sqrt{s} = 350$ GeV (left panel) 
and $500$ GeV (right panel).}
\end{figure}

\begin{figure}[ht]
\centering
$\begin{array}{cc}
\hspace*{-4.5cm}
d\sigma/dM_{ZZ} [\textrm{fb}]
& 
\hspace*{-4.5cm}
d\sigma/dM_{ZZ} [\textrm{fb}] \\
\includegraphics[width=7.7cm, height=7.5cm]
{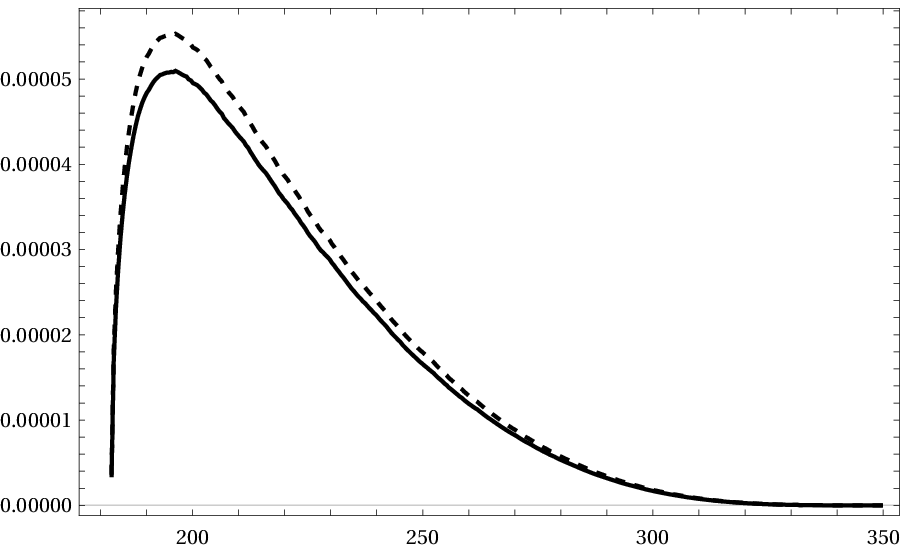}
&
\includegraphics[width=7.7cm, height=7.5cm]
{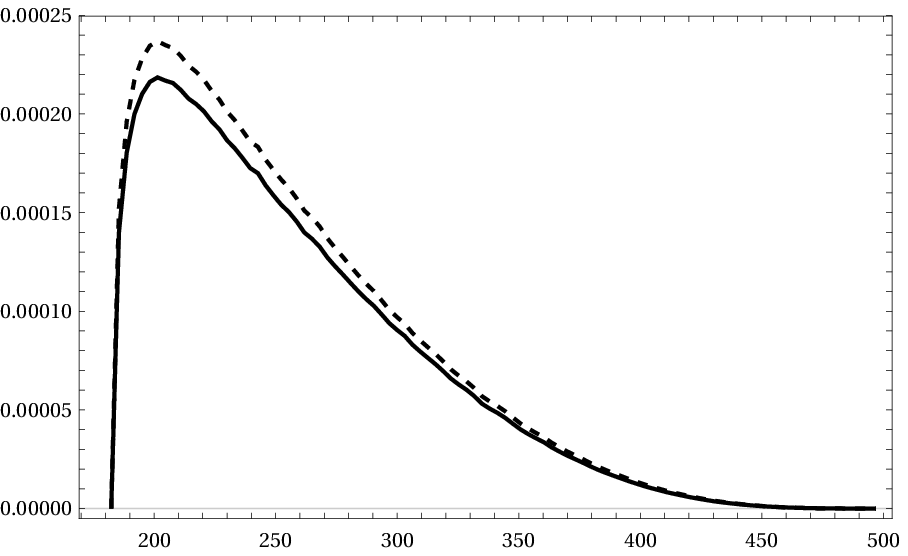}
\\
\hspace{6cm}
M_{ZZ} [\textrm{GeV}]  &
\hspace{6cm}
M_{ZZ} [\textrm{GeV}]
\end{array}$
\caption{\label{ZZ} Differential cross sections
with respect to $M_{ZZ}$ for process 
$e^- e^- \rightarrow e^- e^- H^* 
\rightarrow e^- e^- (ZZ)$ at $\sqrt{s} = 350$ 
GeV (left panel), $500$ GeV (right panel).}
\end{figure}

In Fig.~\ref{gammagamma}, we plot  the 
differential cross section for processes 
$e^-e^- \rightarrow e^-e^- \gamma\gamma$ 
with respect to $M_{\gamma\gamma}$ 
at CMS energies $\sqrt{s} = 350$ GeV (left panel) 
and at $500$ GeV (right panel). We find a peak of 
on-shell $H\rightarrow \gamma\gamma$ 
at $M_{\gamma\gamma}\sim 125$ GeV.
In Fig.~\ref{Zgamma}, the 
differential cross sections for the processes 
$e^-e^- \rightarrow e^-e^- Z\gamma$ with 
respect to $M_{Z\gamma}$ 
at CMS energies $\sqrt{s} = 350$ GeV (left panel) 
and at $500$ GeV (right panel) are shown. 
We also find a peak of 
on-shell $H\rightarrow Z\gamma$ 
at $M_{Z\gamma}\sim 125$ GeV. 
In Fig.~\ref{ZZ}, differential cross sections
with respect to $M_{ZZ}$ for process 
$e^- e^- \rightarrow e^- e^- H^* 
\rightarrow e^- e^- (ZZ)$ at $\sqrt{s} = 350$ 
GeV (left panel), $500$ GeV (right panel).
In these Figures, the 
tree-level cross sections are plotted with 
solid line, and the cross sections with 
full one-loop electroweak 
corrections to $H^* \rightarrow ZZ$ are
shown with dashed line. We find that 
the one-loop off-shell effects 
are visible and they must be taken into 
account at future colliders.
\section{Conclusions}
We have presented one-loop on-shell and 
off-shell decays 
$H\rightarrow VV$ 
with $VV=\gamma\gamma, Z\gamma, ZZ$ in this 
paper. The effects of one-loop on-shell and 
off-shell Higgs decays in Higgs productions 
at $e^-e^-$ collision have then studied. 
We find 
that the impacts of one-loop Higgs decays 
$H\rightarrow VV$
are significant and they are must be taken 
into account at $e^-e^-$ collision.\\

{\bf Acknowledgment:}~
This research is funded by Vietnam National
University, Ho Chi Minh City (VNU-HCM) 
under grant number C$2022$-$18$-$14$.\\ 
\section*{Appendix: Feynman diagrams}
\begin{figure}[ht]
\centering
\includegraphics[width=15cm, height=3cm]
{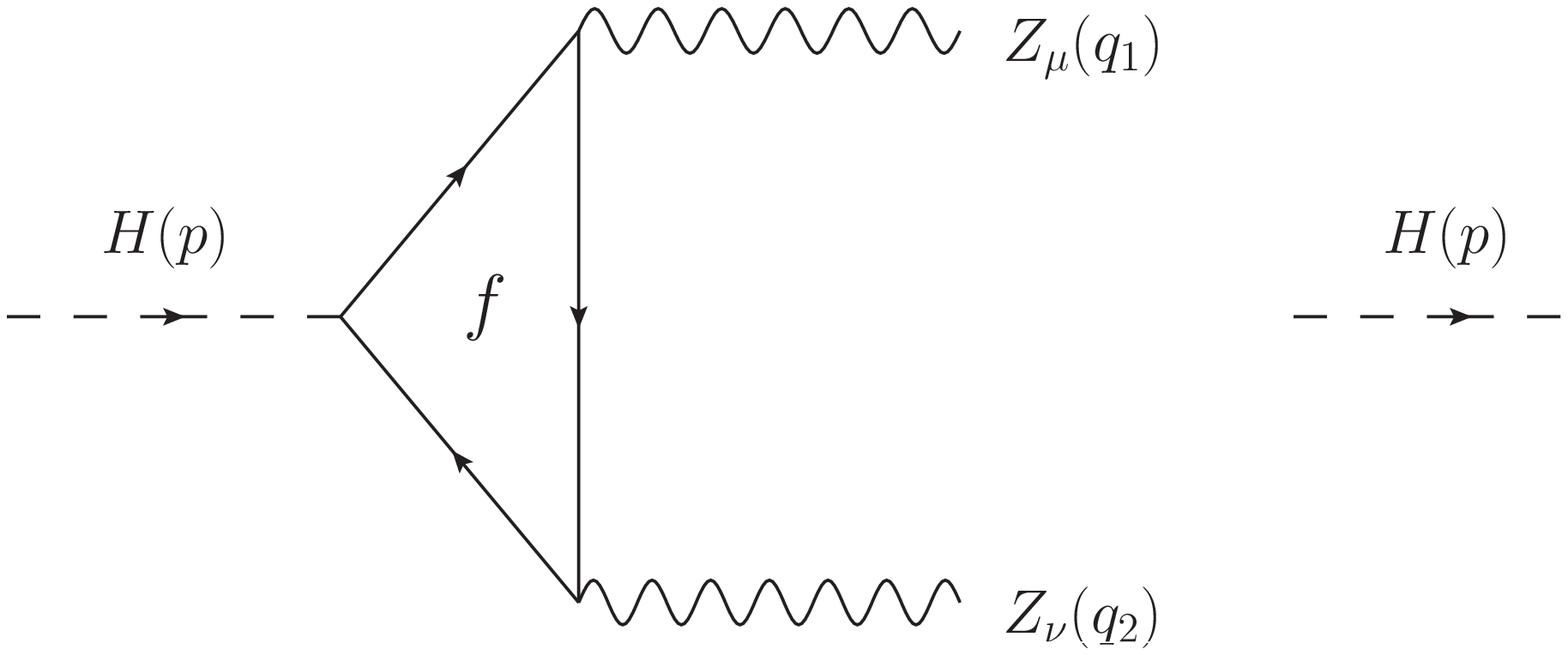}
\caption{one-loop Feynman
diagrams with exchanging $f$ in the loop (Group $1$).}
\end{figure}
\begin{figure}[ht]
\centering
\includegraphics[width=15.0cm, height=3.2cm]
{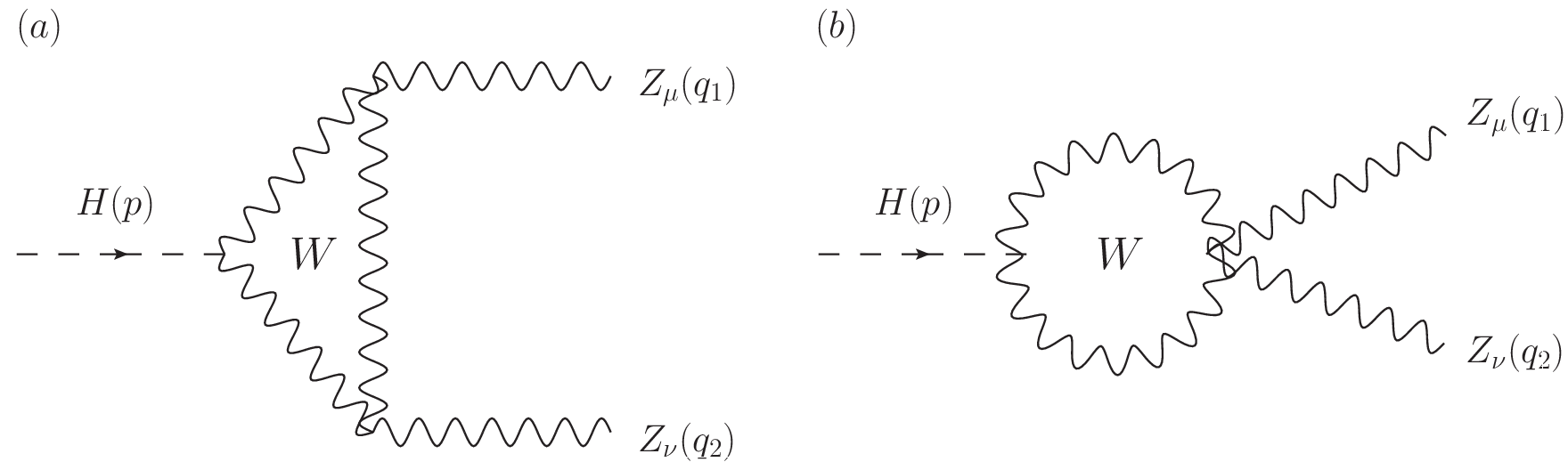}
\includegraphics[width=15.0cm, height=3.0cm]
{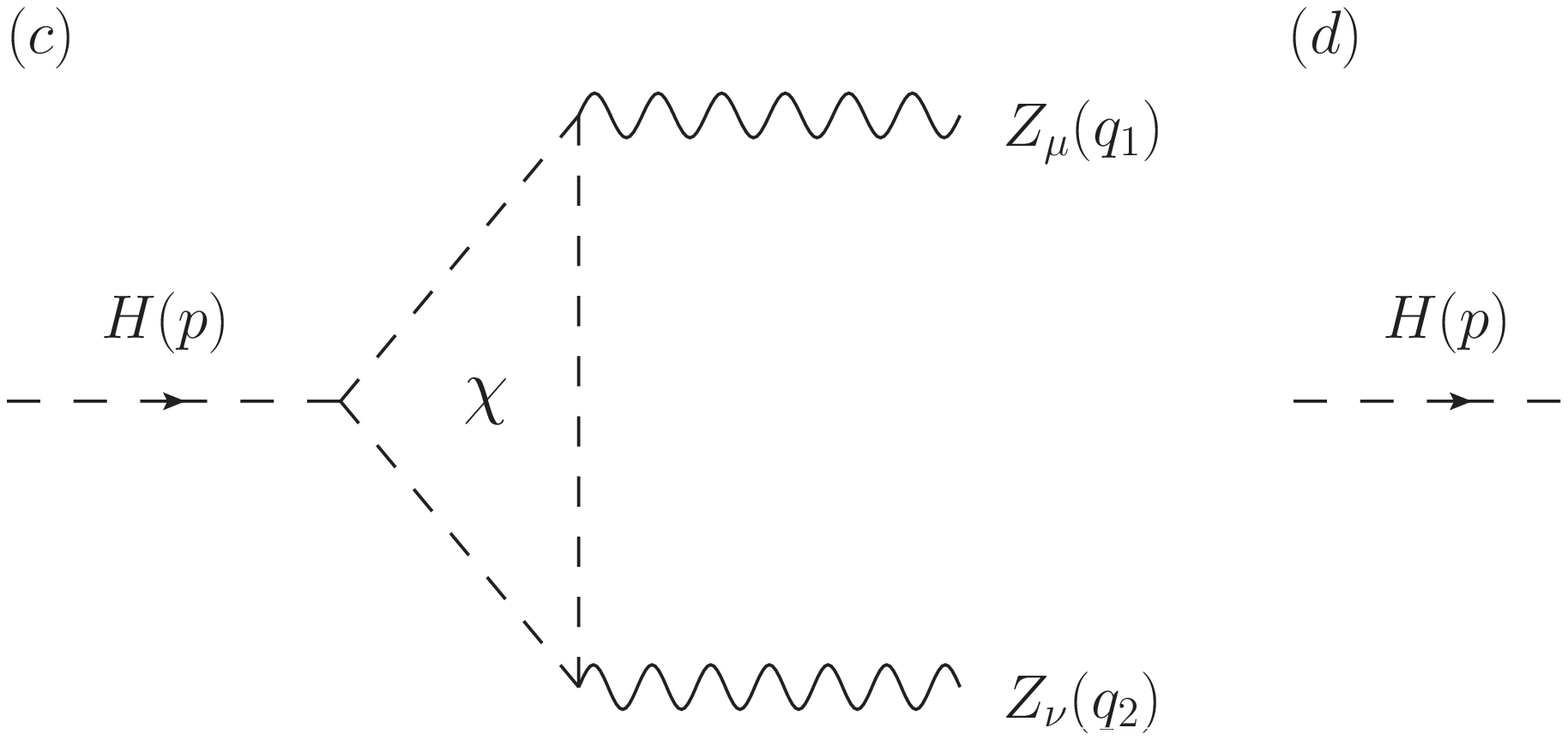}
\includegraphics[width=15.0cm, height=3.0cm]
{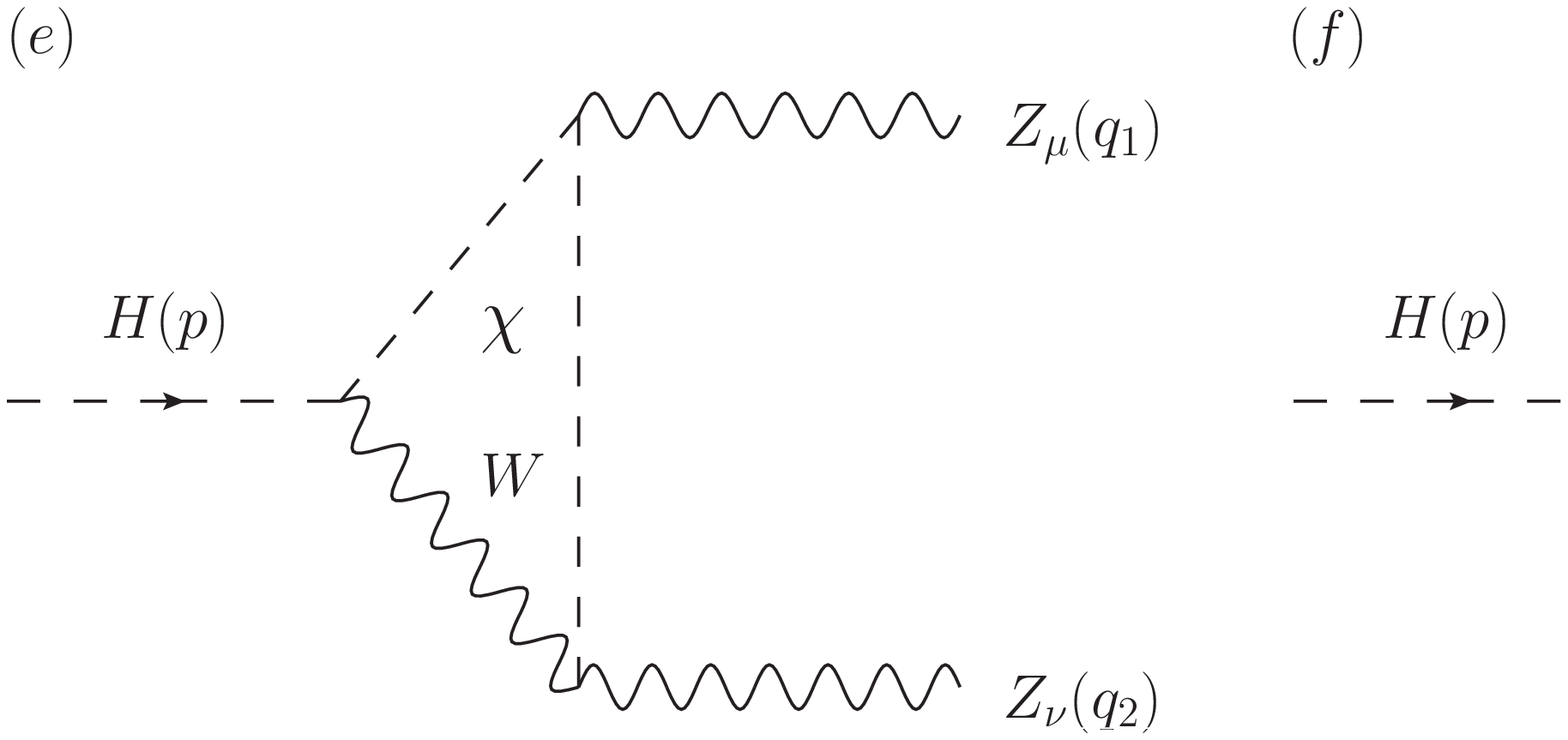}
\includegraphics[width=15.0cm, height=3.0cm]
{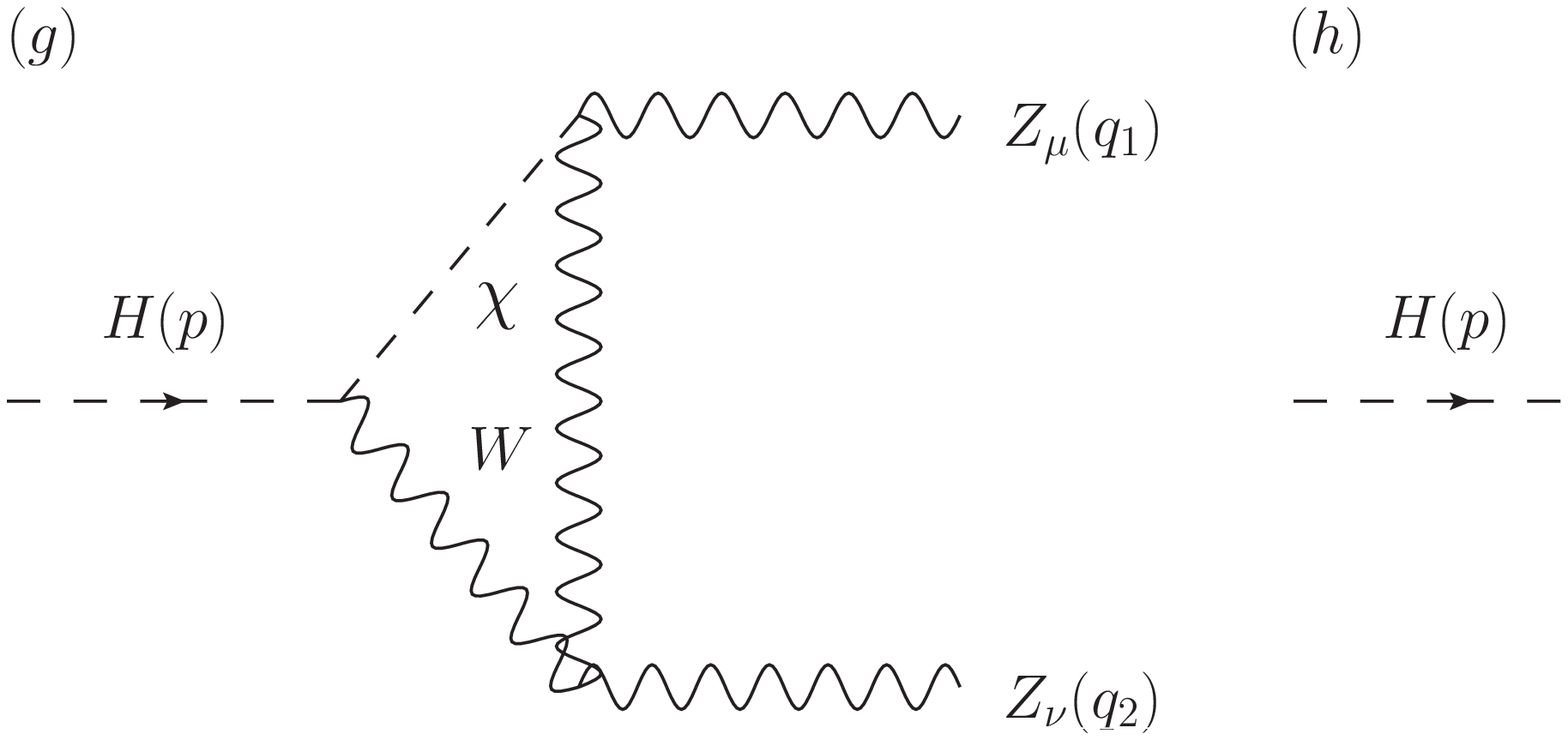}
\includegraphics[width=15.0cm, height=3.0cm]
{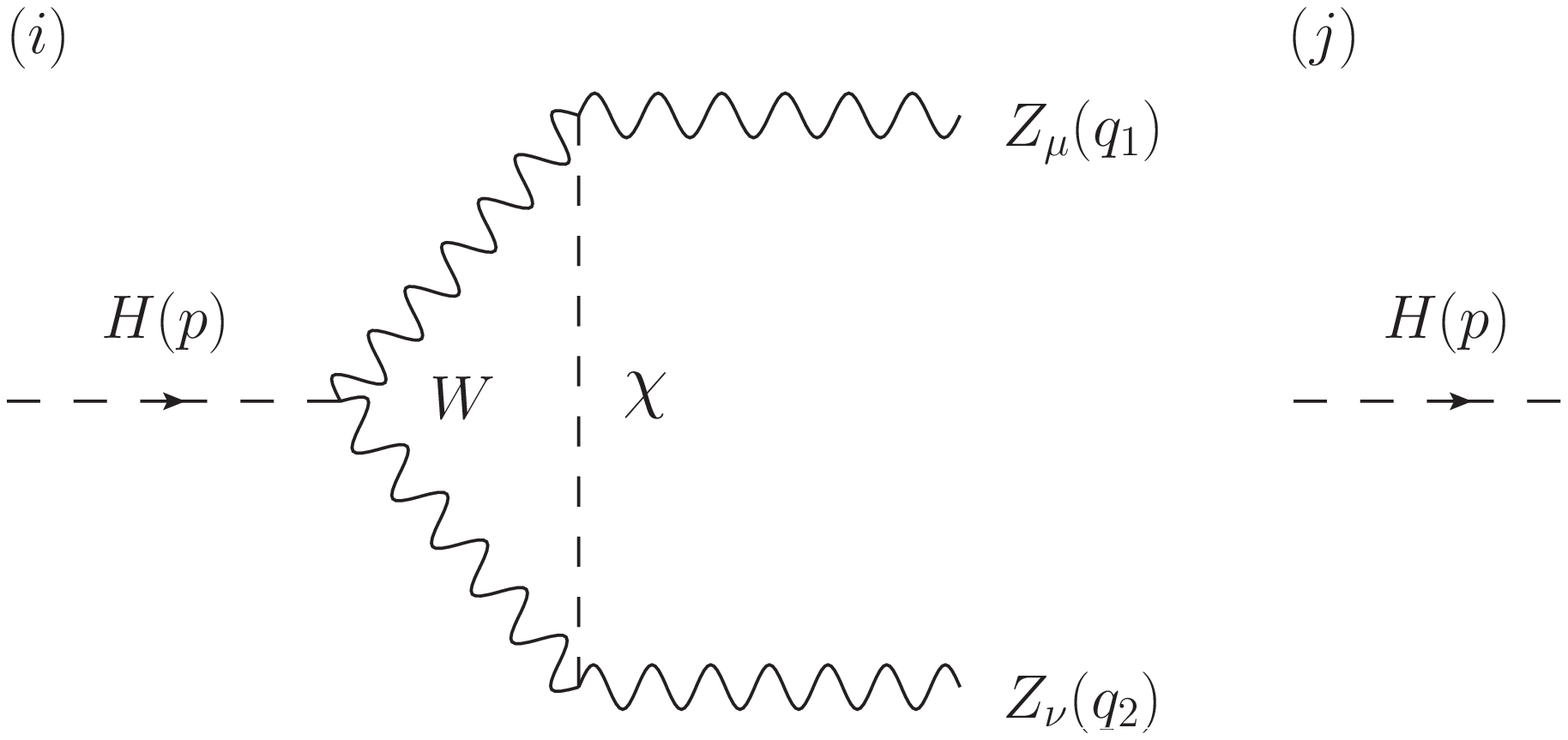}
\includegraphics[width=15.0cm, height=2.8cm]
{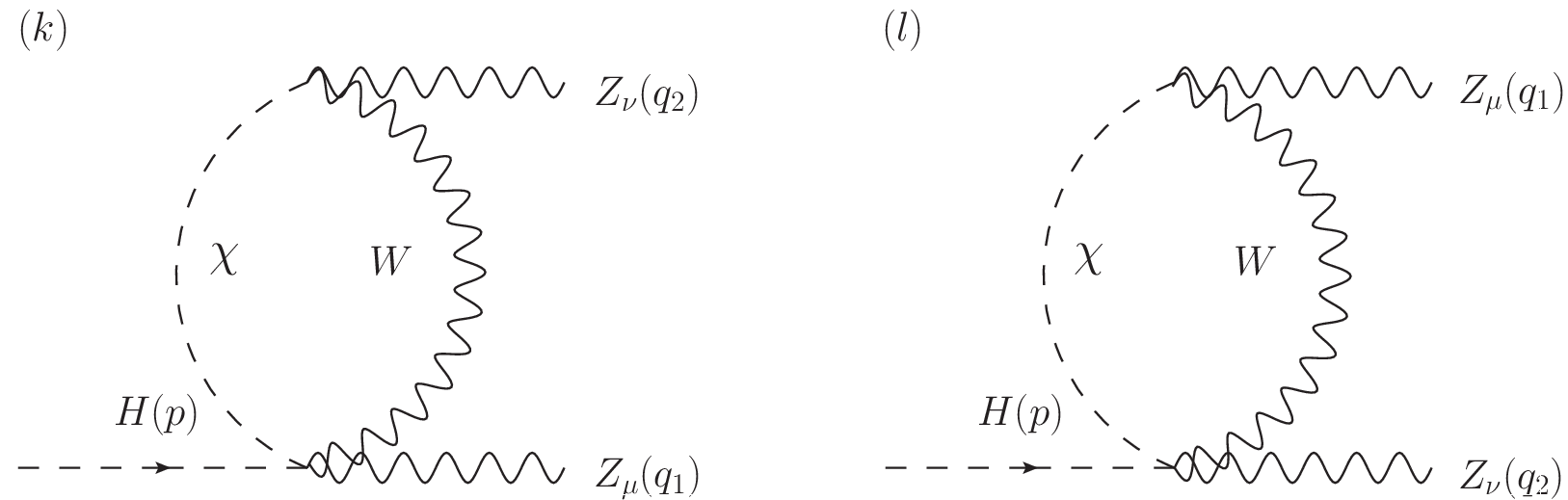}
\includegraphics[width=15.0cm, height=3.0cm]
{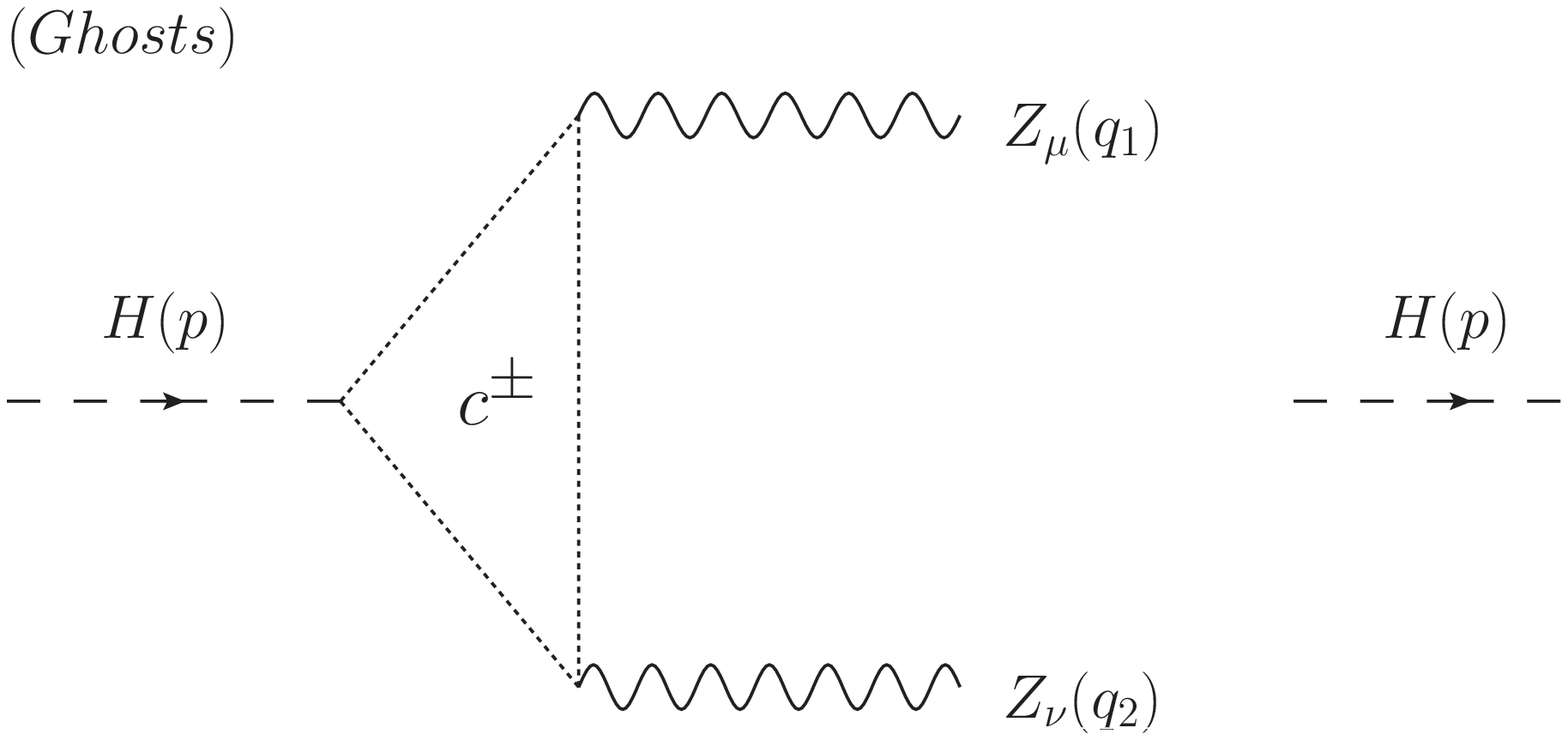}
\caption{one-loop Feynman
diagrams with exchanging $W, \chi$ 
and ghost particles
in the loop (Group $2$).}
\end{figure}
\begin{figure}[ht]
\centering
\includegraphics[width=15.0cm, height=3.0cm]
{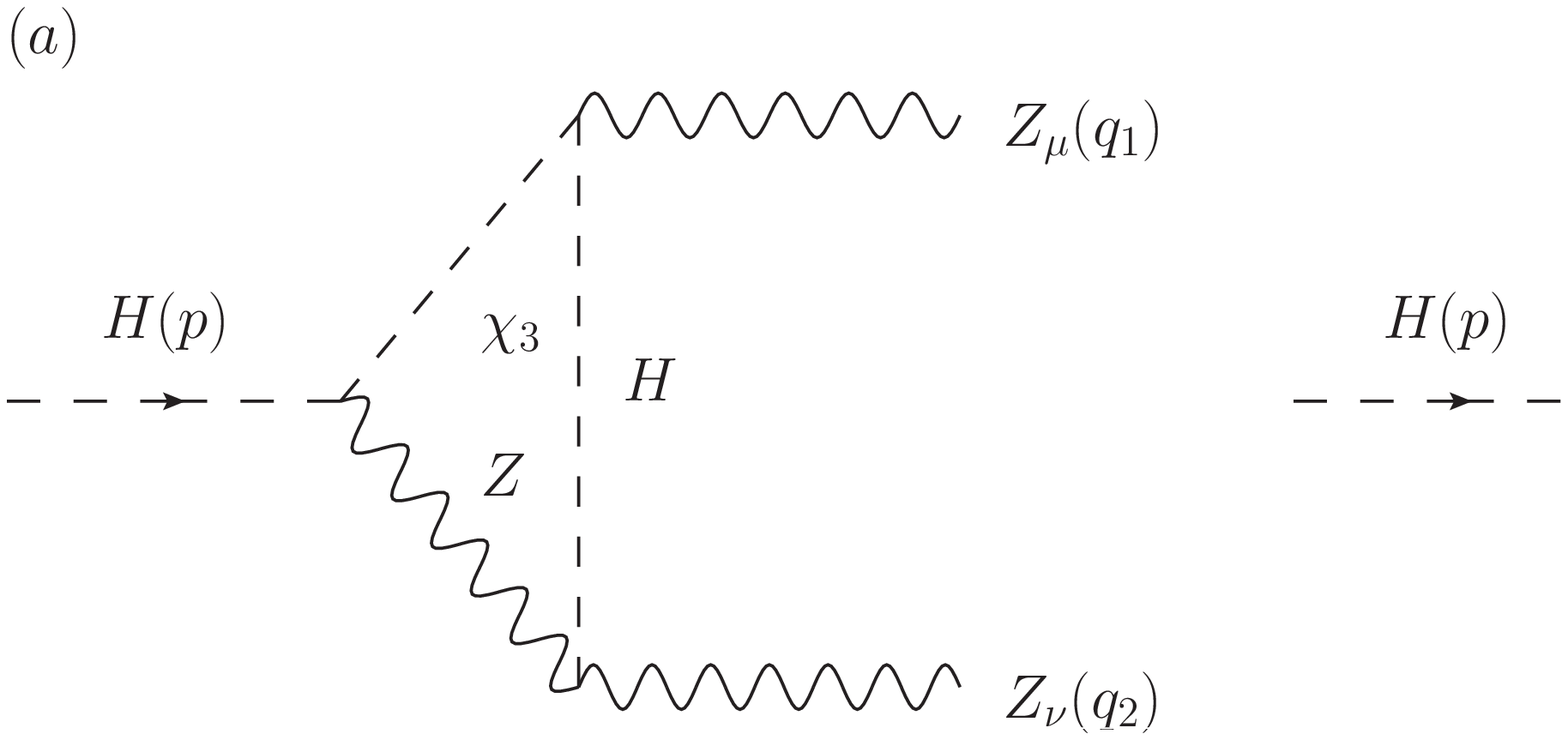}
\includegraphics[width=15.0cm, height=3.0cm]
{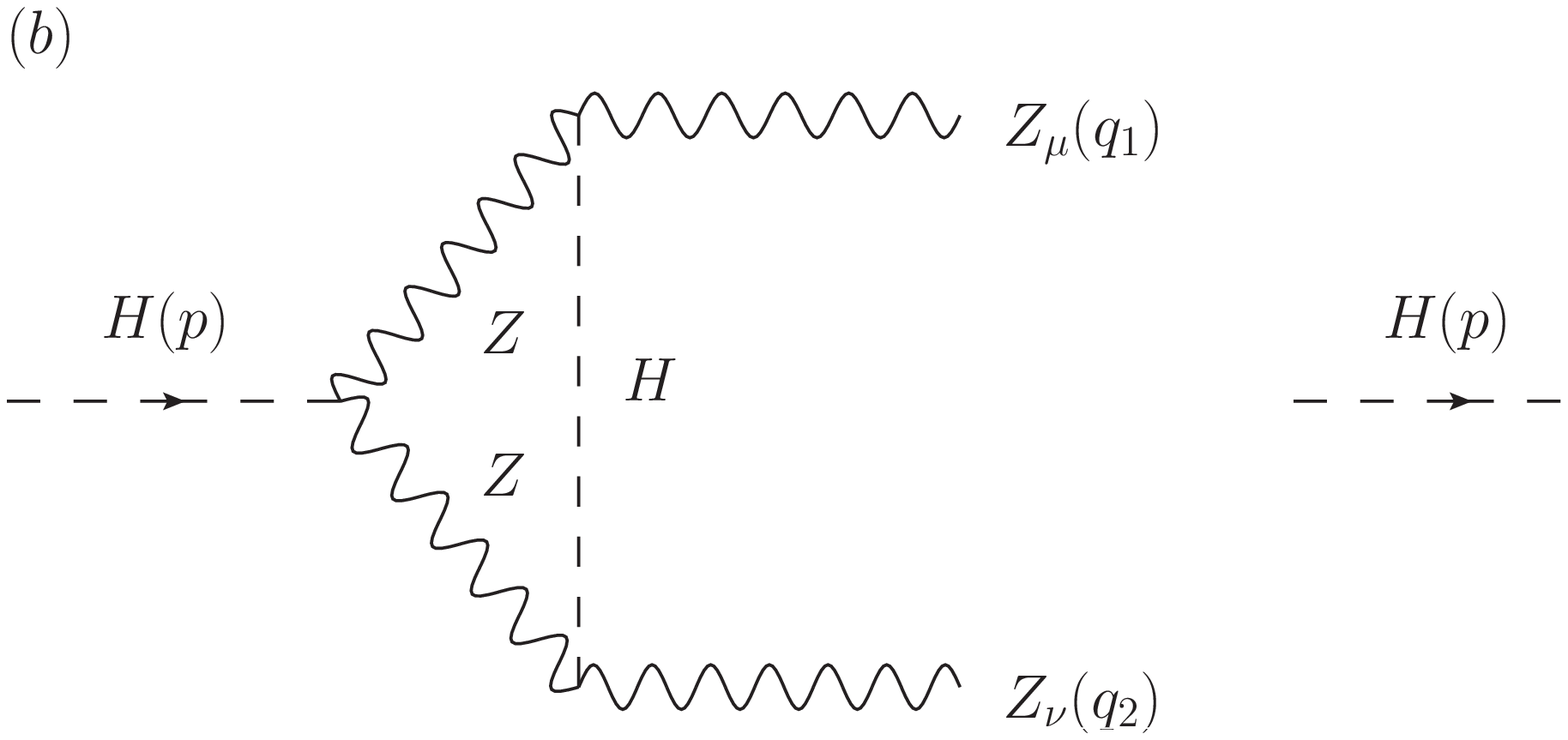}
\includegraphics[width=15.0cm, height=3.0cm]
{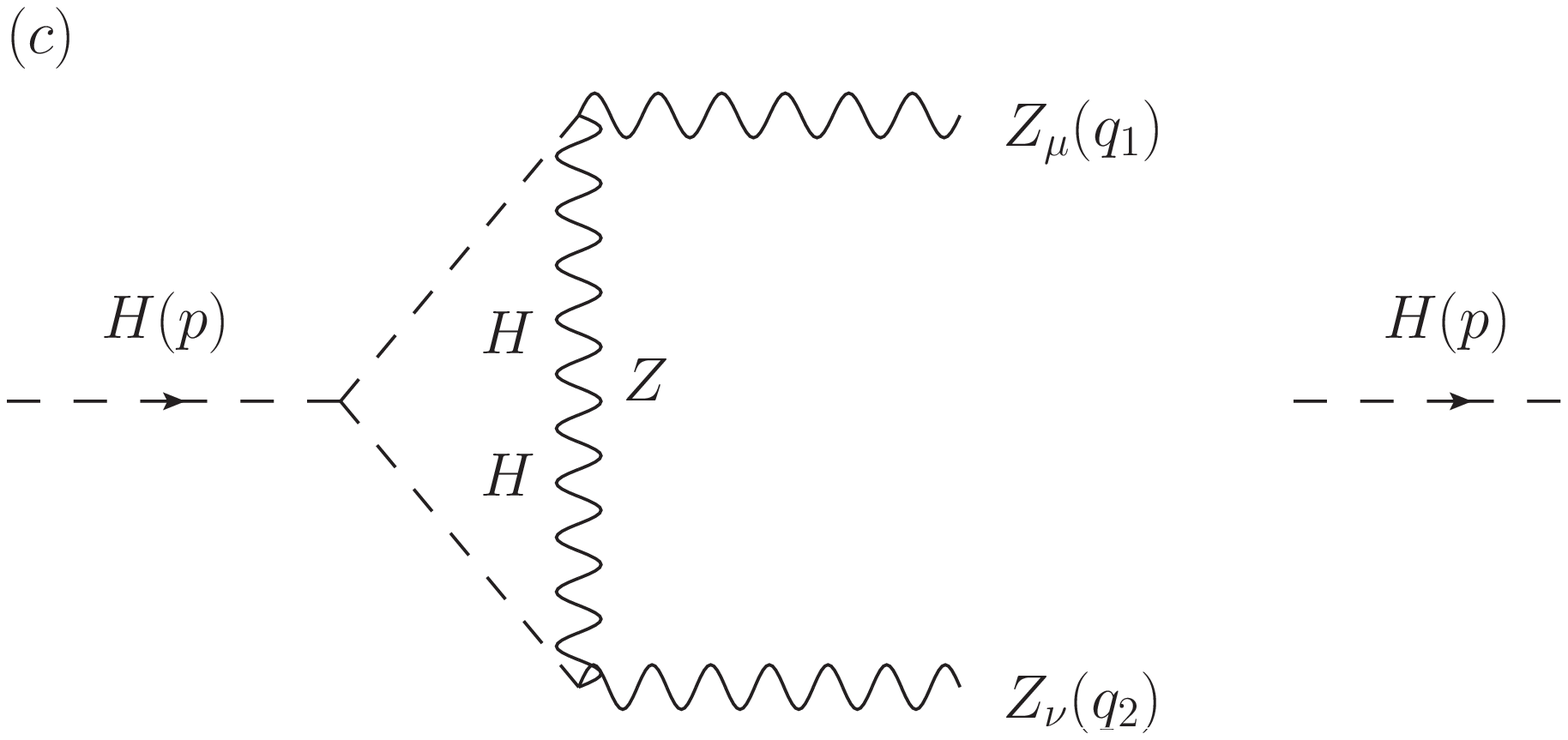}
\includegraphics[width=15.0cm, height=3.3cm]
{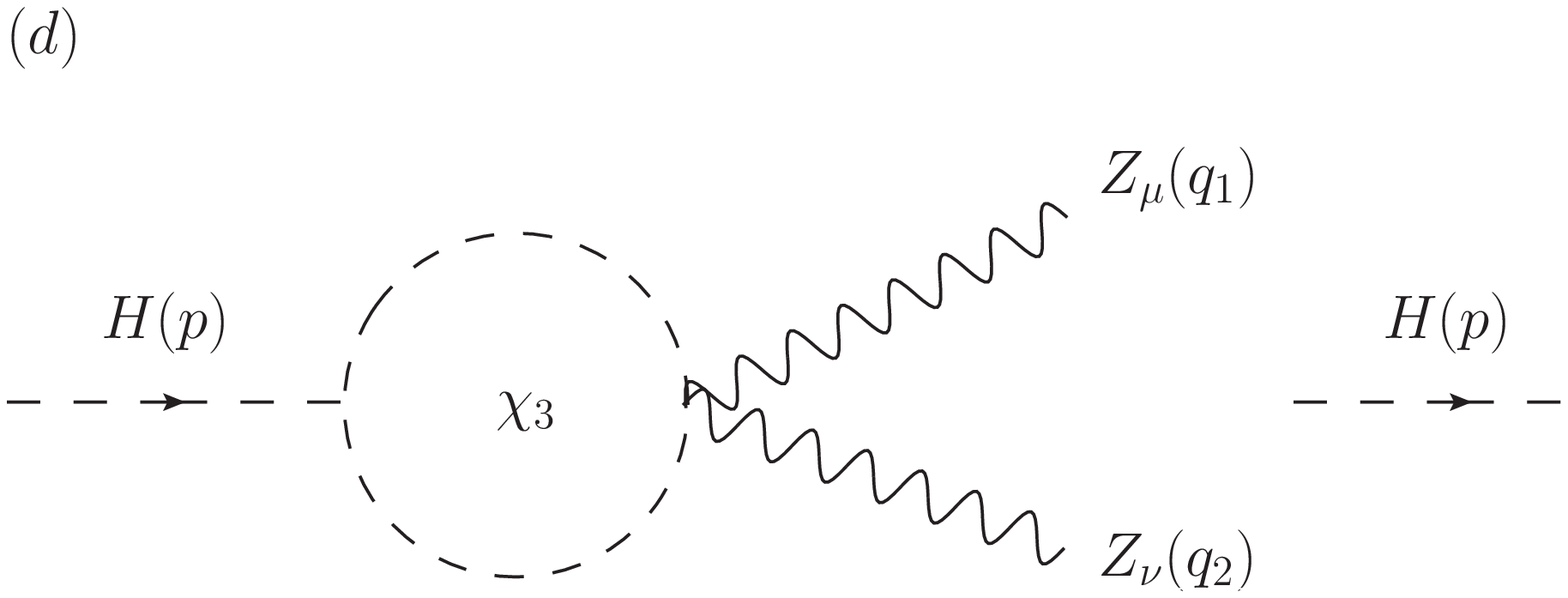}
\includegraphics[width=15.0cm, height=2.6cm]
{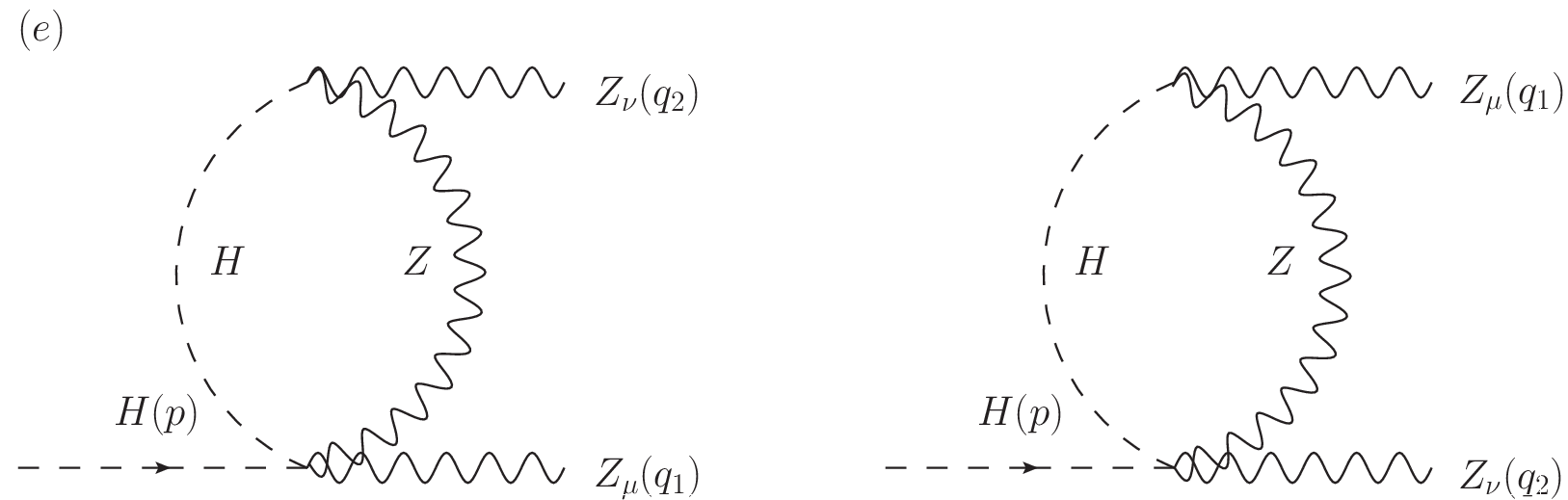}
\caption{one-loop Feynman
diagrams with exchanging $Z, \chi_3$ 
and $H$
in the loop (Group $3$).}
\end{figure}
\begin{figure}[ht]
\centering
\includegraphics[width=6cm, height=3cm]
{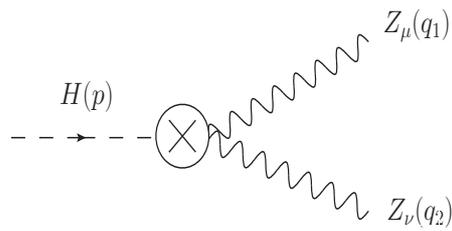}
\caption{Group $0$: counter-term Feynman
diagram.}
\end{figure}
\section*{References}

\end{document}